# An algorithm for multiplication of split-octonions

*Aleksandr Cariow*[1], *Galina Cariowa*[1], *Bartosz Kubsik*[1]

[1]*Faculty of Computer Sciences and Information Technologies, Żołnierska 52,*
*71-210 Szczecin, Poland*
*{atariov, gtariova, bkubsik}@wi.zut.edu.pl*
*tel. +48 91 4495573*

**Abstract:**

*In this paper we introduce efficient algorithm for the multiplication of split-octonions. The direct multiplication of two split-octonions requires 64 real multiplications and 56 real additions. More effective solutions still do not exist. We show how to compute a product of the split-octonions with 28 real multiplications and 92 real additions. During synthesis of the discussed algorithm we use the fact that product of two split-octonions may be represented as vector-matrix product. The matrix that participates in the product calculating has unique structural properties that allow performing its advantageous decomposition. Namely this decomposition leads to significant reducing of the multiplicative complexity of split-octonions multiplication.*

## 1. Introduction

The Clifford and hypercomplex algebras [1] are seeing increased application to digital signal and image processing [2-5], computer graphics and machine vision [6-7], telecommunications [8-10] and in public key cryptography [11]. Among other arithmetical operations in the Clifford and hypercomplex algebras, multiplication is the most time consuming one. The reason for this is, because the usual multiplication of these numbers requires $N(N-1)$ real additions and $N^2$ real multiplication. It is easy to see that the increasing of dimension of hypernumber increases the computational complexity of the multiplication. Therefore, reducing the computational complexity of the multiplication of Clifford and hypercomplex numbers is an important theoretical and practical task. Efficient algorithms for the multiplication of various hypercomplex numbers already exist [12-21]. No such algorithms for the multiplication of the split-octonions have been proposed. In this paper, an efficient algorithm for this purpose is suggested.

## 2. Formulation of the problem

A split-octonions is defined as follows:

$$\breve{o} = b_0 + b_1 e_1 + b_2 e_2 + b_3 e_3 + b_4 e_4 + b_5 e_5 + b_6 e_6 + b_7 e_7$$

where $\{b_i\}, i = 0,1,...,7$ are real numbers, and $\{e_j\}, j = 1,2,...,7$ are imaginary units whose products are defined by the following table [23]:

| $\times$ | $1$ | $e_1$ | $e_2$ | $e_3$ | $e_4$ | $e_5$ | $e_6$ | $e_7$ |
|---|---|---|---|---|---|---|---|---|
| $1$ | $1$ | $e_1$ | $e_2$ | $e_3$ | $e_4$ | $e_5$ | $e_6$ | $e_7$ |
| $e_1$ | $e_1$ | $-1$ | $e_3$ | $-e_2$ | $-e_5$ | $e_4$ | $-e_7$ | $e_6$ |
| $e_2$ | $e_2$ | $-e_3$ | $-1$ | $e_1$ | $-e_6$ | $e_7$ | $e_4$ | $-e_5$ |
| $e_3$ | $e_3$ | $e_2$ | $-e_1$ | $-1$ | $-e_7$ | $-e_6$ | $e_5$ | $e_4$ |
| $e_4$ | $e_4$ | $e_5$ | $e_6$ | $e_7$ | $1$ | $e_1$ | $e_2$ | $e_3$ |
| $e_5$ | $e_5$ | $-e_4$ | $-e_7$ | $e_6$ | $-e_1$ | $1$ | $e_3$ | $-e_2$ |
| $e_6$ | $e_6$ | $e_7$ | $-e_4$ | $-e_5$ | $-e_2$ | $-e_3$ | $1$ | $e_1$ |
| $e_7$ | $e_7$ | $-e_6$ | $e_5$ | $-e_4$ | $-e_3$ | $e_2$ | $-e_1$ | $1$ |

Suppose we must to compute the product of two split-octonions $\breve{o}_3 = \breve{o}_1 \breve{o}_2$ ,

where

$$\breve{o}_1 = x_0 + x_1 e_1 + x_2 e_2 + x_3 e_3 + x_4 e_4 + x_5 e_5 + x_6 e_6 + x_7 e_7 \,,$$

$$\breve{o}_2 = b_0 + b_1 e_1 + b_2 e_2 + b_3 e_3 + b_4 e_4 + b_5 e_5 + b_6 e_6 + b_7 e_7 \,,$$

$$\breve{o}_3 = y_0 + y_1 e_1 + y_2 e_2 + y_3 e_3 + y_4 e_4 + y_5 e_5 + y_6 e_6 + y_7 e_7 \,.$$

Using “pen and paper” method we can write:

$$\begin{aligned}
\breve{o}_3 = {} & x_0 b_0 + x_0 b_1 e_1 + x_0 b_2 e_2 + x_0 b_3 e_3 + x_0 b_4 e_4 + x_0 b_5 e_5 + x_0 b_6 e_6 + x_0 b_7 e_7 \\
& + x_1 b_0 e_1 + x_1 b_1 e_1 e_1 + x_1 b_2 e_1 e_2 + x_1 b_3 e_1 e_3 + x_1 b_4 e_1 e_4 + x_1 b_5 e_1 e_5 + x_1 b_6 e_1 e_6 + x_1 b_7 e_1 e_7 \\
& + x_2 b_0 e_2 + x_2 b_1 e_2 e_1 + x_2 b_2 e_2 e_2 + x_2 b_3 e_2 e_3 + x_2 b_4 e_2 e_4 + x_2 b_5 e_2 e_5 + x_2 b_6 e_2 e_6 + x_2 b_7 e_2 e_7 \\
& + x_3 b_0 e_3 + x_3 b_1 e_3 e_1 + x_3 b_2 e_3 e_2 + x_3 b_3 e_3 e_3 + x_3 b_4 e_3 e_4 + x_3 b_5 e_3 e_5 + x_3 b_6 e_3 e_6 + x_3 b_7 e_3 e_7 \\
& + x_4 b_0 e_4 + x_4 b_1 e_4 e_1 + x_4 b_2 e_4 e_2 + x_4 b_3 e_4 e_3 + x_4 b_4 e_4 e_4 + x_4 b_5 e_4 e_5 + x_4 b_6 e_4 e_6 + x_4 b_7 e_4 e_7 \\
& + x_5 b_0 e_5 + x_5 b_1 e_5 e_1 + x_5 b_2 e_5 e_2 + x_5 b_3 e_5 e_3 + x_5 b_4 e_5 e_4 + x_5 b_5 e_5 e_5 + x_5 b_6 e_5 e_6 + x_5 b_7 e_5 e_7 \\
& + x_6 b_0 e_6 + x_6 b_1 e_6 e_1 + x_6 b_2 e_6 e_2 + x_6 b_3 e_6 e_3 + x_6 b_4 e_6 e_4 + x_6 b_5 e_6 e_5 + x_6 b_6 e_6 e_6 + x_6 b_7 e_6 e_7 \\
& + x_7 b_0 e_7 + x_7 b_1 e_7 e_1 + x_7 b_2 e_7 e_2 + x_7 b_3 e_7 e_3 + x_7 b_4 e_7 e_4 + x_7 b_5 e_7 e_5 + x_7 b_6 e_7 e_6 + x_7 b_7 e_7 e_7
\end{aligned}$$

Then we have:

$$\begin{aligned}
y_0 &= x_0 b_0 - x_1 b_1 - x_2 b_2 - x_3 b_3 + x_4 b_4 + x_5 b_5 + x_6 b_6 + x_7 b_7 \,, \\
y_1 &= x_0 b_1 + x_1 b_0 + x_2 b_3 - x_3 b_2 + x_4 b_5 - x_5 b_4 + x_6 b_7 - x_7 b_6 \,, \\
y_2 &= x_0 b_2 - x_1 b_3 + x_2 b_0 + x_3 b_1 + x_4 b_6 - x_5 b_7 - x_6 b_4 + x_7 b_5 \,, \\
y_3 &= x_0 b_3 + x_1 b_2 - x_2 b_1 + x_3 b_0 + x_4 b_7 + x_5 b_6 - x_6 b_5 - x_7 b_4 \,, \\
y_4 &= x_0 b_4 + x_1 b_5 + x_2 b_6 + x_3 b_7 + x_4 b_0 - x_5 b_1 - x_6 b_2 - x_7 b_3 \,, \\
y_5 &= x_0 b_5 - x_1 b_4 - x_2 b_7 + x_3 b_6 + x_4 b_1 + x_5 b_0 - x_6 b_3 + x_7 b_2 \,, \\
y_6 &= x_0 b_6 + x_1 b_7 - x_2 b_4 - x_3 b_5 + x_4 b_2 + x_5 b_3 + x_6 b_0 - x_7 b_1 \,, \\
y_7 &= x_0 b_7 - x_1 b_6 + x_2 b_5 - x_3 b_4 + x_4 b_3 - x_5 b_2 + x_6 b_1 + x_7 b_0 \,.
\end{aligned}$$

We can see that the schoolbook method of multiplication of two split-octonions requires 64 real multiplications and 56 real additions.

Using the matrix notation, we can rewrite the above relations as follows:

$$\mathbf{Y}_{8\times1} = \mathbf{B}_8 \mathbf{X}_{8\times1} \qquad (1)$$

where

$$\mathbf{X}_{8\times1} = [x_0, x_1, x_2, x_3, x_4, x_5, x_6, x_7]^{\mathrm{T}} , \ \mathbf{Y}_{8\times1} = [y_0, y_1, y_2, y_3 . y_4, y_5, y_6, y_7]^{\mathrm{T}} ,$$

$$\mathbf{B}_8 = \left[\begin{array}{cccc|cccc}
b_0 & -b_1 & -b_2 & -b_3 & b_4 & b_5 & b_6 & b_7 \\
b_1 & b_0 & b_3 & -b_2 & b_5 & -b_4 & b_7 & -b_6 \\
b_2 & -b_3 & b_0 & b_1 & b_6 & -b_7 & -b_4 & b_5 \\
b_3 & b_2 & -b_1 & b_0 & b_7 & b_6 & -b_5 & -b_4 \\
\hline
b_4 & b_5 & b_6 & b_7 & b_0 & -b_1 & -b_2 & -b_3 \\
b_5 & -b_4 & -b_7 & b_6 & b_1 & b_0 & -b_3 & b_2 \\
b_6 & b_7 & -b_4 & -b_5 & b_2 & b_3 & b_0 & -b_1 \\
b_7 & -b_6 & b_5 & -b_4 & b_3 & -b_2 & b_1 & b_0
\end{array}\right],$$

The direct realization of (1) requires 64 real multiplications and 56 real additions too. We shall present the algorithm, which reduce arithmetical complexity to 28 real multiplications and 92 real additions.

## 3. Synthesis of a rationalized algorithm for multiplying two split-octonions

It easy to see, that the matrix $\mathbf{B}_8$ can be represented as an algebraic sum of a symmetric Toeplitz-type matrix and another matrix which has many zero elements $\mathbf{B}_8 = \mathbf{B}_8^{(1)} + 2\mathbf{M}_8^{(0)}$ :

where

$$\mathbf{B}_8^{(1)} = \left[\begin{array}{cccc:cccc} b_0 & -b_1 & -b_2 & -b_3 & b_4 & b_5 & b_6 & b_7 \\ b_1 & b_0 & b_3 & b_2 & b_5 & -b_4 & b_7 & b_6 \\ b_2 & b_3 & b_0 & b_1 & b_6 & b_7 & -b_4 & b_5 \\ b_3 & b_2 & b_1 & b_0 & b_7 & b_6 & b_5 & -b_4 \\ \hdashline b_4 & b_5 & b_6 & b_7 & b_0 & -b_1 & -b_2 & -b_3 \\ b_5 & -b_4 & b_7 & b_6 & b_1 & b_0 & b_3 & b_2 \\ b_6 & b_7 & -b_4 & b_5 & b_2 & b_3 & b_0 & b_1 \\ b_7 & b_6 & b_5 & -b_4 & b_3 & b_2 & b_1 & b_0 \end{array}\right], \quad \mathbf{M}_8^{(0)} = \left[\begin{array}{cccc:cccc} 0 & 0 & 0 & 0 & 0 & 0 & 0 & 0 \\ 0 & 0 & 0 & -b_2 & 0 & 0 & 0 & -b_6 \\ 0 & -b_3 & 0 & 0 & 0 & -b_7 & 0 & 0 \\ 0 & 0 & -b_1 & 0 & 0 & 0 & -b_5 & 0 \\ \hdashline 0 & 0 & 0 & 0 & 0 & 0 & 0 & 0 \\ 0 & 0 & -b_7 & 0 & 0 & 0 & -b_3 & 0 \\ 0 & 0 & 0 & -b_5 & 0 & 0 & 0 & -b_1 \\ 0 & -b_6 & 0 & 0 & 0 & -b_2 & 0 & 0 \end{array}\right].$$

Then we can write

$$\mathbf{Y}_{8\times1} = \mathbf{\Sigma}_{8\times16}(\mathbf{B}_8^{(1)} \oplus 2\mathbf{M}_8^{(0)})\mathbf{P}_{16\times8}\mathbf{X}_{8\times1} \tag{2}$$

where

$$\mathbf{\Sigma}_{8\times16} = \left[\begin{array}{cccccccc:cccccccc} 1 &&&&&&& & 1 &&&&&&& \\ & 1 &&&&&& && 1 &&&&&& \\ && 1 &&&&& &&& 1 &&&&& \\ &&& 1 &&&& &&&& 1 &&&& \\ &&&& 1 &&& &&&&& 1 &&& \\ &&&&& 1 && &&&&&& 1 && \\ &&&&&& 1 & &&&&&&& 1 & \\ &&&&&&& 1 &&&&&&&& 1 \end{array}\right], \quad \mathbf{P}_{16\times8} = \left[\begin{array}{cccccccc} 1 &&&&&&& \\ & 1 &&&&&& \\ && 1 &&&&& \\ &&& 1 &&&& \\ &&&& 1 &&& \\ &&&&& 1 && \\ &&&&&& 1 & \\ &&&&&&& 1 \\ \hdashline 1 &&&&&&& \\ & 1 &&&&&& \\ && 1 &&&&& \\ &&& 1 &&&& \\ &&&& 1 &&& \\ &&&&& 1 && \\ &&&&&& 1 & \\ &&&&&&& 1 \end{array}\right].$$

and sign “ $\oplus$ ” – denotes the direct sum of two matrices [24],

We can see that the matrix $\mathbf{B}_8^{(1)}$ has a unique block structure:

$$\mathbf{B}_8^{(1)} = \left[\begin{array}{c:c} \mathbf{A}_4 & \mathbf{B}_4 \\ \hdashline \mathbf{B}_4 & \mathbf{A}_4 \end{array}\right],$$

where

$$\mathbf{A}_4 = \left[\begin{array}{cc:cc} b_0 & -b_1 & -b_2 & -b_3 \\ b_1 & b_0 & b_3 & b_2 \\ \hdashline b_2 & b_3 & b_0 & b_1 \\ b_3 & b_2 & b_1 & b_0 \end{array}\right], \quad \mathbf{B}_4 = \left[\begin{array}{cc:cc} b_4 & b_5 & b_6 & b_7 \\ b_5 & -b_4 & b_7 & b_6 \\ \hdashline b_6 & b_7 & -b_4 & b_5 \\ b_7 & b_6 & b_5 & -b_4 \end{array}\right].$$

It is easily verify [25-27] that the matrix $\mathbf{B}_8^{(1)}$ with this structure can be factorized, than the computational procedure for multiplication of the split-octonions can be represented as follows:

$$\mathbf{Y}_{8\times1} = \mathbf{\Sigma}_{8\times16}\mathbf{W}_{16}\mathbf{D}_{16}^{(0)}\mathbf{W}_{16}\mathbf{P}_{16\times8}\mathbf{X}_{8\times1} \tag{3}$$

where

$$\mathbf{W}_{16} = (\mathbf{H}_2 \otimes \mathbf{I}_4) \oplus \mathbf{I}_8 = \left[\begin{array}{cccccccc:cccccccc}
1 & & & & 1 & & & & & & & & & & & \\
 & 1 & & & & 1 & & & & & & & & & & \\
 & & 1 & & & & 1 & & & & & & & & & \\
 & & & 1 & & & & 1 & & & & & & & & \\
1 & & & & -1 & & & & & & & & & & & \\
 & 1 & & & & -1 & & & & & & & & & & \\
 & & 1 & & & & -1 & & & & & & & & & \\
 & & & 1 & & & & -1 & & & & & & & & \\
\hdashline
 & & & & & & & & 1 & & & & & & & \\
 & & & & & & & & & 1 & & & & & & \\
 & & & & & & & & & & 1 & & & & & \\
 & & & & & & & & & & & 1 & & & & \\
 & & & & & & & & & & & & 1 & & & \\
 & & & & & & & & & & & & & 1 & & \\
 & & & & & & & & & & & & & & 1 & \\
 & & & & & & & & & & & & & & & 1
\end{array}\right],$$

$$\mathbf{D}_{16}^{(0)} = \frac{1}{2}\left(\left(\mathbf{A}_4 + \mathbf{B}_4\right) \oplus \left(\mathbf{A}_4 - \mathbf{B}_4\right)\right) \oplus 2\mathbf{M}_8^{(0)}$$

$\mathbf{H}_2 = \begin{bmatrix} 1 & 1 \\ 1 & -1 \end{bmatrix}$ - is the order 2 Hadamard matrix, $\mathbf{I}_N$ is the order $N$ identity matrix, and „$\otimes$” – denotes the Kronecker product of two matrices [24].

The matrices $(\mathbf{A}_4 + \mathbf{B}_4)$ and $(\mathbf{A}_4 - \mathbf{B}_4)$ have the following structures:

$$\left(\mathbf{A}_4 + \mathbf{B}_4\right) = \begin{bmatrix} b_0 + b_4 & -b_1 + b_5 & -b_2 + b_6 & -b_3 + b_7 \\ b_1 + b_5 & b_0 - b_4 & b_3 + b_7 & b_2 + b_6 \\ b_2 + b_6 & b_3 + b_7 & b_0 - b_4 & b_1 + b_5 \\ b_3 + b_7 & b_2 + b_6 & b_1 + b_5 & b_0 - b_4 \end{bmatrix} = \mathbf{E}_4^{(0)},$$

$$\left(\mathbf{A}_4 - \mathbf{B}_4\right) = \begin{bmatrix} b_0 - b_4 & -b_1 - b_5 & -b_2 - b_6 & -b_3 - b_7 \\ b_1 - b_5 & b_0 + b_4 & b_3 - b_7 & b_2 - b_6 \\ b_2 - b_6 & b_3 - b_7 & b_0 + b_4 & b_1 - b_5 \\ b_3 - b_7 & b_2 - b_6 & b_1 - b_5 & b_0 + b_4 \end{bmatrix} = \mathbf{F}_4^{(0)}$$

Fig. 1 shows a data flow diagram of the rationalized algorithm for computation of a product of a split-octonions. In this paper, data flow diagrams are oriented from left to right. Straight lines in the figures denote the operations of data transfer. Points where lines converge denote summation. The dashed lines indicate the sign change operation. We deliberately use the usual lines without arrows on purpose, so as not to clutter the picture. The rectangles indicate the matrix–vector multiplications with the matrix inscribed inside a rectangle.

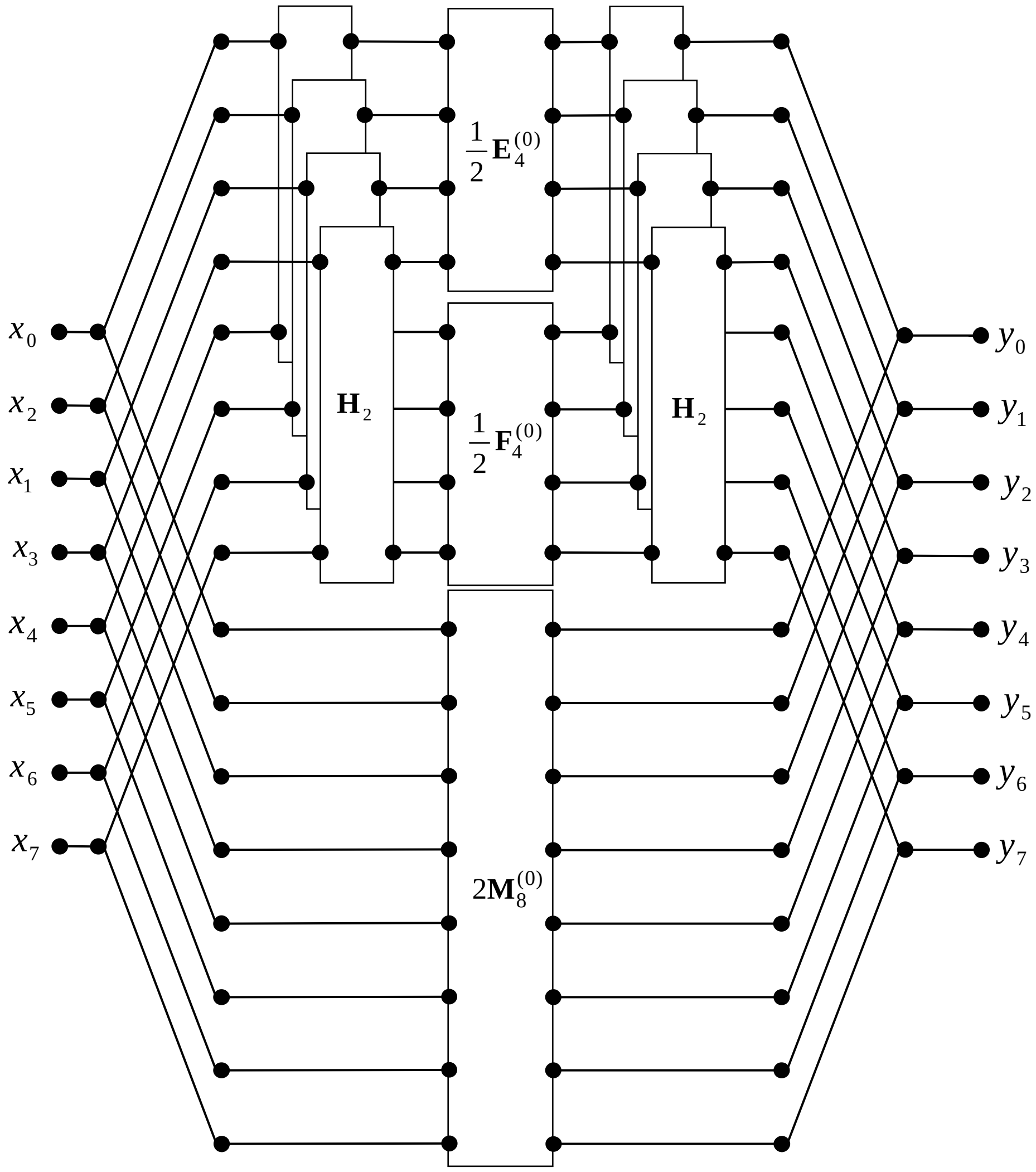

**Fig. 1.** Data flow diagram for rationalized split-octonion multiplication algorithm in accordance with the procedure (3).

Let us consider the structures of the matrices $\mathbf{E}_4^{(0)}$ and $\mathbf{F}_4^{(0)}$. The matrix $\mathbf{E}_4^{(0)}$ can be decomposed as an algebraic sum of a symmetric Toeplitz matrix and another matrix which has many zero elements $\mathbf{E}_4^{(0)} = \mathbf{E}_4^{(1)} + 2\mathbf{M}_4^{(1)}$:

$$\mathbf{E}_4^{(1)} = \left[\begin{array}{cc|cc} b_0 - b_4 & b_1 + b_5 & b_2 + b_6 & b_3 + b_7 \\ b_1 + b_5 & b_0 - b_4 & b_3 + b_7 & b_2 + b_6 \\ \hline b_2 + b_6 & b_3 + b_7 & b_0 - b_4 & b_1 + b_5 \\ b_3 + b_7 & b_2 + b_6 & b_1 + b_5 & b_0 - b_4 \end{array}\right] = \left[\begin{array}{c|c} \mathbf{A}_2 & \mathbf{B}_2 \\ \hline \mathbf{B}_2 & \mathbf{A}_2 \end{array}\right], \; \mathbf{M}_4^{(1)} = \begin{bmatrix} b_4 & -b_1 & -b_2 & -b_3 \\ 0 & 0 & 0 & 0 \\ 0 & 0 & 0 & 0 \\ 0 & 0 & 0 & 0 \end{bmatrix}$$

The matrix $\mathbf{F}_4^{(0)}$ can be also represented as an algebraic sum of a symmetric Toeplitz matrix and another matrix which has many zero elements $\mathbf{F}_4^{(0)} = \mathbf{F}_4^{(1)} + 2\mathbf{M}_4^{(2)}$ :

$$\mathbf{F}_4^{(1)} = \left[\begin{array}{cc|cc} b_0+b_4 & b_1-b_5 & b_2-b_6 & b_3-b_7 \\ b_1-b_5 & b_0+b_4 & b_3-b_7 & b_2-b_6 \\ \hline b_2-b_6 & b_3-b_7 & b_0+b_4 & b_1-b_5 \\ b_3-b_7 & b_2-b_6 & b_1-b_5 & b_0+b_4 \end{array}\right] = \left[\begin{array}{c|c} \mathbf{C}_2 & \mathbf{D}_2 \\ \hline \mathbf{D}_2 & \mathbf{C}_2 \end{array}\right],$$

$$\mathbf{M}_4^{(2)} = \begin{bmatrix} -b_4 & -b_1 & -b_2 & -b_3 \\ 0 & 0 & 0 & 0 \\ 0 & 0 & 0 & 0 \\ 0 & 0 & 0 & 0 \end{bmatrix}$$

It is easily to verify [25-27] that the matrices $\mathbf{E}_4^{(1)}$ and $\mathbf{F}_4^{(1)}$ can be factorized as:

$$\mathbf{E}_4^{(1)} = (\mathbf{H}_2 \otimes \mathbf{I}_2)\frac{1}{2}[(\mathbf{A}_2 + \mathbf{B}_2) \oplus (\mathbf{A}_2 - \mathbf{B}_2)](\mathbf{H}_2 \otimes \mathbf{I}_2) \tag{4}$$

$$\mathbf{F}_4^{(1)} = (\mathbf{H}_2 \otimes \mathbf{I}_2)\frac{1}{2}[(\mathbf{C}_2 + \mathbf{D}_2) \oplus (\mathbf{C}_2 - \mathbf{D}_2)](\mathbf{H}_2 \otimes \mathbf{I}_2) \tag{5}$$

$$\mathbf{A}_2 = \left[\begin{array}{c|c} b_0-b_4 & b_1+b_5 \\ \hline b_1+b_5 & b_0-b_4 \end{array}\right], \ \mathbf{B}_2 = \left[\begin{array}{c|c} b_2+b_6 & b_3+b_7 \\ \hline b_3+b_7 & b_2+b_6 \end{array}\right],$$

$$\mathbf{C}_2 = \left[\begin{array}{c|c} b_0+b_4 & b_1-b_5 \\ \hline b_1-b_5 & b_0+b_4 \end{array}\right], \ \mathbf{D}_2 = \left[\begin{array}{c|c} b_2-b_6 & b_3-b_7 \\ \hline b_3-b_7 & b_2-b_6 \end{array}\right],$$

$$(\mathbf{A}_2 + \mathbf{B}_2) = \left[\begin{array}{c|c} b_0-b_4+b_2+b_6 & b_1+b_5+b_3+b_7 \\ \hline b_1+b_5+b_3+b_7 & b_0-b_4+b_2+b_6 \end{array}\right] = \mathbf{E}_2^{(0)},$$

$$(\mathbf{A}_2 - \mathbf{B}_2) = \left[\begin{array}{c|c} b_0-b_4-b_2-b_6 & b_1+b_5-b_3-b_7 \\ \hline b_1+b_5-b_3-b_7 & b_0-b_4-b_2-b_6 \end{array}\right] = \mathbf{F}_2^{(0)},$$

$$(\mathbf{C}_2 + \mathbf{D}_2) = \left[\begin{array}{c|c} b_0+b_4+b_2-b_6 & b_1-b_5+b_3-b_7 \\ \hline b_1-b_5+b_3-b_7 & b_0+b_4+b_2-b_6 \end{array}\right] = \mathbf{K}_2^{(0)},$$

$$(\mathbf{C}_2 - \mathbf{D}_2) = \left[\begin{array}{c|c} b_0+b_4-b_2+b_6 & b_1-b_5-b_3+b_7 \\ \hline b_1-b_5-b_3+b_7 & b_0+b_4-b_2+b_6 \end{array}\right] = \mathbf{L}_2^{(0)},$$

Substituting (4) and (5) in (3) we can write:

$$\mathbf{Y}_{8\times1} = \mathbf{\Sigma}_{8\times16}\mathbf{W}_{16}\mathbf{\Sigma}_{16\times24}\mathbf{W}_{24}^{(1)}\mathbf{D}_{24}^{(1)}\mathbf{W}_{24}^{(1)}\mathbf{P}_{24\times16}\mathbf{W}_{16}\mathbf{P}_{16\times8}\mathbf{X}_{8\times1} \tag{6}$$

where

$$\mathbf{D}_{24}^{(1)} = \mathbf{Q}_8^{(1)} \oplus \mathbf{Q}_8^{(2)} \oplus \mathbf{M}_8^{(0)},$$

$$\mathbf{Q}_8^{(1)} = \frac{1}{4}(\mathbf{E}_2^{(0)} \oplus \mathbf{F}_2^{(0)}) \oplus \mathbf{T}_4^{(0)}, \ \mathbf{Q}_8^{(2)} = \frac{1}{4}(\mathbf{K}_2^{(0)} \oplus \mathbf{L}_2^{(0)}) \oplus \mathbf{T}_4^{(1)}, \ \mathbf{T}_4^{(0)} = \mathbf{T}_4^{(1)} = diag(b_4, b_1, b_2, b_3),$$

$$
\mathbf{W}_{24}^{(1)}=\left[\begin{array}{c|c|c}
\begin{array}{c|c}
\begin{array}{cc|cc}
1 & & 1 & \\
 & 1 & & 1 \\ \hline
1 & & -1 & \\
 & 1 & & -1
\end{array} & \mathbf{0}_4 \\ \hline
\mathbf{0}_4 & \begin{array}{cccc} 1 & & & \\ & 1 & & \\ & & 1 & \\ & & & 1 \end{array}
\end{array} & \mathbf{0}_8 & \mathbf{0}_8 \\ \hline
\mathbf{0}_8 & \begin{array}{c|c}
\begin{array}{cc|cc}
1 & & 1 & \\
 & 1 & & 1 \\ \hline
1 & & -1 & \\
 & 1 & & -1
\end{array} & \mathbf{0}_4 \\ \hline
\mathbf{0}_4 & \begin{array}{cccc} 1 & & & \\ & 1 & & \\ & & 1 & \\ & & & 1 \end{array}
\end{array} & \mathbf{0}_8 \\ \hline
\mathbf{0}_8 & \mathbf{0}_8 & \begin{array}{cccccccc} 1 & & & & & & & \\ & 1 & & & & & & \\ & & 1 & & & & & \\ & & & 1 & & & & \\ & & & & 1 & & & \\ & & & & & 1 & & \\ & & & & & & 1 & \\ & & & & & & & 1 \end{array}
\end{array}\right],
$$

$$
\mathbf{\Sigma}_{16\times 24}=\left[\begin{array}{cccc|cccc|cccc|cccc|cccccccc}
1 & & & & 1 & -1 & -1 & -1 \\
 & 1 \\
 & & 1 \\
 & & & 1 \\ \hline
 & & & & & & & & 1 & & & & -1 & -1 & -1 & -1 \\
 & & & & & & & & & 1 \\
 & & & & & & & & & & 1 \\
 & & & & & & & & & & & 1 \\ \hline
 & & & & & & & & & & & & & & & & 1 \\
 & & & & & & & & & & & & & & & & & 1 \\
 & & & & & & & & & & & & & & & & & & 1 \\
 & & & & & & & & & & & & & & & & & & & 1 \\
 & & & & & & & & & & & & & & & & & & & & 1 \\
 & & & & & & & & & & & & & & & & & & & & & 1 \\
 & & & & & & & & & & & & & & & & & & & & & & 1 \\
 & & & & & & & & & & & & & & & & & & & & & & & 1
\end{array}\right],
$$

$$
\mathbf{P}_{24\times 16} = \left[\begin{array}{cccc|cccc|cccccccc}
1 &&&&& &&&&& &&&&& \\
&1 &&&& &&&&& &&&&& \\
&&1 &&& &&&&& &&&&& \\
&&&1 && &&&&& &&&&& \\
\hline
1 &&&&& &&&&& &&&&& \\
&1 &&&& &&&&& &&&&& \\
&&1 &&& &&&&& &&&&& \\
&&&1 && &&&&& &&&&& \\
\hline
&&&& 1 &&&&& &&&&& & \\
&&&&& 1 &&&& &&&&& & \\
&&&&& & 1 &&& &&&&& & \\
&&&&& && 1 && &&&&& & \\
\hline
&&&& 1 &&&&& &&&&& & \\
&&&&& 1 &&&& &&&&& & \\
&&&&& & 1 &&& &&&&& & \\
&&&&& && 1 && &&&&& & \\
\hline
&&&&& &&& 1 & &&&&& & \\
&&&&& &&&& 1 &&&&& & \\
&&&&& &&&&& 1 &&&& & \\
&&&&& &&&&& & 1 &&& & \\
&&&&& &&&&& && 1 && & \\
&&&&& &&&&& &&& 1 & & \\
&&&&& &&&&& &&&& 1 & \\
&&&&& &&&&& &&&&& 1
\end{array}\right],
$$

Fig. 2 shows a data flow diagram of the rationalized algorithm for multiplying of two split-octonions at the second stage of synthesis. The circles in this figure show the operation of multiplication by a variable (or constant) inscribed inside a circle.

Consider now the matrices $\mathbf{E}_2^{(0)}$, $\mathbf{F}_2^{(0)}$, $\mathbf{K}_2^{(0)}$, and $\mathbf{L}_2^{(0)}$. As can be seen, these matrices also have a "good" structures leading to a decrease in the number of real multiplications during calculation of the split-octonion product.

$$
\mathbf{E}_2^{(0)} = \left[\begin{array}{c|c} b_0-b_4+b_2+b_6 & b_1+b_5+b_3+b_7 \\ \hline b_1+b_5+b_3+b_7 & b_0-b_4+b_2+b_6 \end{array}\right] = \left[\begin{array}{c|c} a & b \\ \hline b & a \end{array}\right] = \mathbf{H}_2 \frac{1}{2}[(a+b)\oplus(a-b)]\mathbf{H}_2 ,
$$

$$
\mathbf{F}_2^{(0)} = \left[\begin{array}{c|c} b_0-b_4-b_2-b_6 & b_1+b_5-b_3-b_7 \\ \hline b_1+b_5-b_3-b_7 & b_0-b_4-b_2-b_6 \end{array}\right] = \left[\begin{array}{c|c} c & d \\ \hline d & c \end{array}\right] = \mathbf{H}_2 \frac{1}{2}[(c+d)\oplus(c-d)]\mathbf{H}_2 ,
$$

$$
\mathbf{K}_2^{(0)} = \left[\begin{array}{c|c} b_0+b_4+b_2-b_6 & b_1-b_5+b_3-b_7 \\ \hline b_1-b_5+b_3-b_7 & b_0+b_4+b_2-b_6 \end{array}\right] = \left[\begin{array}{c|c} e & f \\ \hline f & e \end{array}\right] = \mathbf{H}_2 \frac{1}{2}[(e+f)\oplus(e-f)]\mathbf{H}_2 ,
$$

$$
\mathbf{L}_2^{(0)} = \left[\begin{array}{c|c} b_0+b_4-b_2+b_6 & b_1-b_5-b_3+b_7 \\ \hline -b_1+b_5+b_3-b_7 & b_0+b_4-b_2+b_6 \end{array}\right] = \left[\begin{array}{c|c} g & h \\ \hline h & g \end{array}\right] = \mathbf{H}_2 \frac{1}{2}[(g+h)\oplus(g-h)]\mathbf{H}_2
$$

Introduce the following notation:

$$
a+b=c_0=b_0-b_4+b_2+b_6+b_1+b_5+b_3+b_7 ,\; a-b=c_1=b_0-b_4+b_2+b_6-b_1-b_5-b_3-b_7 ,
$$

$$
c+d=c_2=b_0-b_4-b_2-b_6+b_1+b_5-b_3-b_7 ,\; c-d=c_3=b_0-b_4-b_2-b_6-b_1-b_5+b_3+b_7 ,
$$

$$e+f=c_4=b_0+b_4+b_2-b_6+b_1-b_5+b_3-b_7\,,\quad e-f=c_5=b_0+b_4+b_2-b_6-b_1+b_5-b_3+b_7\,,$$

$$g+h=c_6=b_0+b_4-b_2+b_6+b_1-b_5-b_3+b_7\,,\quad g-h=c_7=b_0+b_4-b_2+b_6-b_1+b_5+b_3-b_7\,.$$

and

$$s_0=\frac{1}{8}c_0\,,\ s_1=\frac{1}{8}c_1\,,\ s_2=\frac{1}{8}c_2\,,\ s_3=\frac{1}{8}c_3\,,\ s_4=\frac{1}{8}c_4\,,\ s_5=\frac{1}{8}c_5\,,\ s_6=\frac{1}{8}c_6\,,\ s_7=\frac{1}{8}c_7\,.$$

Using the above notations and combining partial decompositions in a single computational procedure we finally can write following:

$$\mathbf{Y}_{8\times1}=\mathbf{\Sigma}_{8\times16}\mathbf{W}_{16}\mathbf{\Sigma}_{16\times24}\mathbf{W}_{24}^{(1)}\mathbf{W}_{24}^{(2)}\mathbf{D}_{24}^{(2)}\mathbf{W}_{24}^{(2)}\mathbf{W}_{24}^{(1)}\mathbf{P}_{24\times16}\mathbf{W}_{16}\mathbf{P}_{16\times8}\mathbf{X}_{8\times1} \tag{7}$$

$$\mathbf{D}_{24}^{(2)}=\mathbf{Q}_8^{(3)}\oplus\mathbf{Q}_8^{(4)}\oplus2\mathbf{M}_8^{(0)}\,,\ \mathbf{Q}_8^{(3)}=s_0\oplus s_1\oplus s_2\oplus s_3\oplus\mathbf{T}_4^{(0)}\,,\ \mathbf{Q}_8^{(4)}=s_4\oplus s_5\oplus s_6\oplus s_7\oplus\mathbf{T}_4^{(1)}\,.$$

$$\mathbf{Q}_8^{(3)}=\left[\begin{array}{c|c}\begin{array}{c|c}\begin{matrix}s_0&0\\0&s_1\end{matrix}&\mathbf{0}_2\\\hline\mathbf{0}_2&\begin{matrix}s_2&0\\0&s_3\end{matrix}\end{array}&\mathbf{0}_4\\\hline\mathbf{0}_4&\mathbf{T}_4^{(0)}\end{array}\right],\quad \mathbf{Q}_8^{(4)}=\left[\begin{array}{c|c}\begin{array}{c|c}\begin{matrix}s_4&0\\0&s_5\end{matrix}&\mathbf{0}_2\\\hline\mathbf{0}_2&\begin{matrix}s_6&0\\0&s_7\end{matrix}\end{array}&\mathbf{0}_4\\\hline\mathbf{0}_4&\mathbf{T}_4^{(1)}\end{array}\right],$$

$$\mathbf{W}_{24}^{(2)}=\left[\begin{array}{c|c|c|c|c|c}
\begin{matrix}1&1\\1&-1\end{matrix}&&&&&\\\hline
&\begin{matrix}1&1\\1&-1\end{matrix}&&&&\\\hline
&&\begin{matrix}1&&&\\&1&&\\&&1&\\&&&1\end{matrix}&&&\\\hline
&&&\begin{matrix}1&1\\1&-1\end{matrix}&&\\\hline
&&&&\begin{matrix}1&1\\1&-1\end{matrix}&\\\hline
&&&&&\begin{matrix}1\\&1\\&&1\\&&&1\\&&&&1\\&&&&&1\\&&&&&&1\\&&&&&&&1\\&&&&&&&&1\\&&&&&&&&&1\\&&&&&&&&&&1\\&&&&&&&&&&&1\end{matrix}
\end{array}\right],$$

**Fig. 3** shows a data flow diagram of the rationalized algorithm for multiplying of two split-octonions at the final stage of the algorithm derivation.

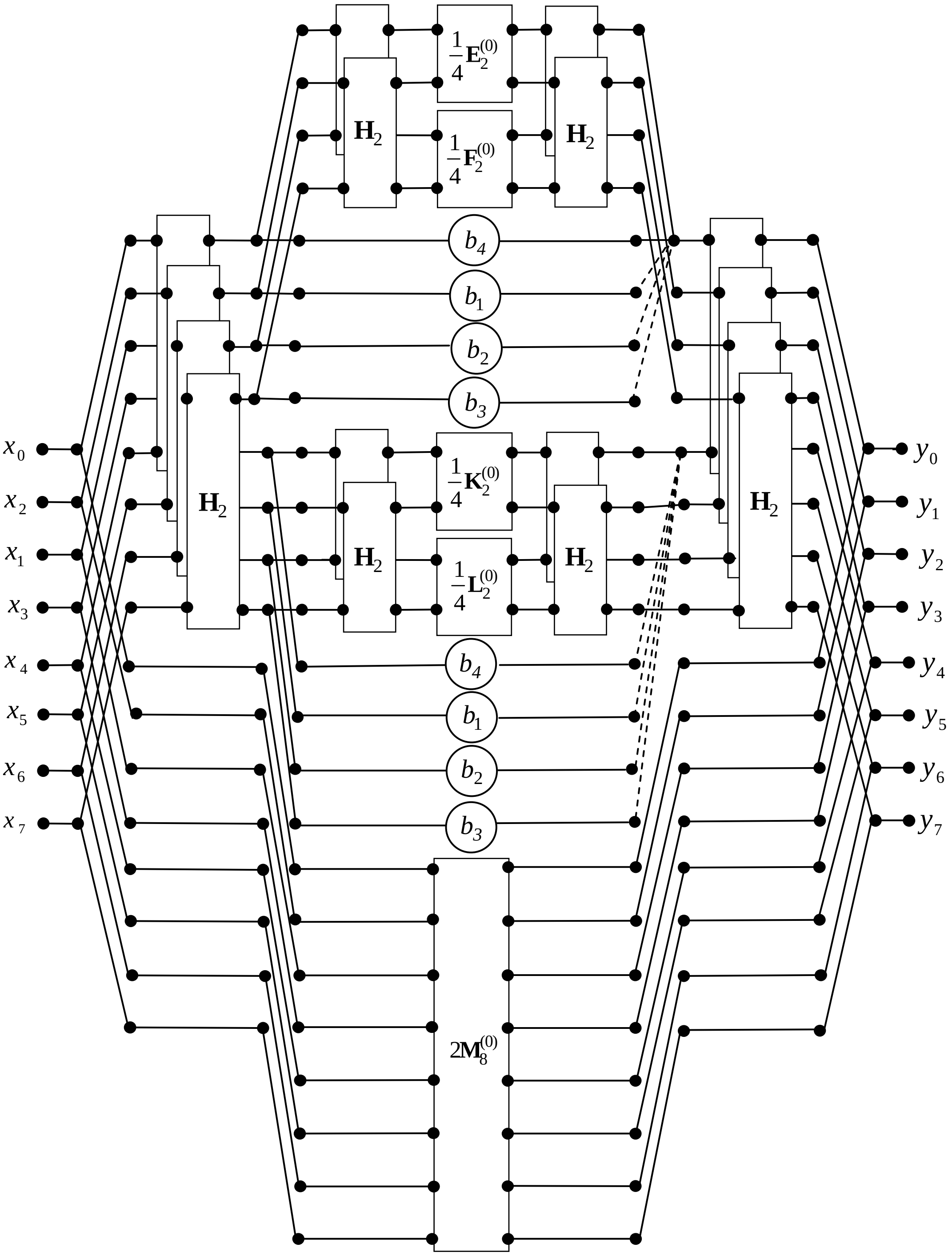

**Fig. 2.** Data flow diagram for rationalized split-octonion multiplication algorithm in accordance with the procedure (6).

We can see that the ordinary approach to calculation of elements $\{s_k\}, k = 0,1,...,7$ requires 56 additions. It is easy to see that the relations for calculation of $\{s_k\}$ contain repeated algebraic sums. Therefore, the number of additions necessary to calculate these elements can be reduced.

Let us first introduce the following notation:

$$z_0 = b_0 - b_4 ,\ z_1 = b_0 + b_4 ,\ z_2 = b_2 + b_6 ,\ z_3 = b_1 + b_5 ,\ z_4 = b_3 + b_7 ,\ z_5 = b_2 - b_6 ,\ z_6 = b_1 - b_5 ,\ z_7 = b_3 - b_7 ,$$

Therefore, we can write:

$$c_0 = z_0 + z_2 + z_3 + z_4 ,\ c_1 = z_0 + z_2 - z_3 - z_4 ,\ c_2 = z_0 - z_2 + z_3 - z_4 ,\ c_3 = z_0 - z_2 - z_3 + z_4 ,$$

$$c_4 = z_1 + z_5 + z_6 + z_7 ,\ c_5 = z_1 + z_5 - z_6 - z_7 ,\ c_6 = z_1 - z_5 + z_6 - z_7 ,\ c_7 = z_1 - z_5 - z_6 + z_7 .$$

Secondly, we introduce the following notation:

$$v_0 = z_0 + z_2 ,\ v_1 = z_0 - z_2 ,\ v_2 = z_1 + z_5 ,\ v_3 = z_1 - z_5 ,\ v_4 = z_3 + z_4 ,\ v_5 = z_3 - z_4 ,\ v_6 = z_6 + z_7 ,\ v_7 = z_6 - z_7 .$$

Then we can write:

$$c_0 = v_0 + v_4 ,\ c_1 = v_0 - v_4 ,\ c_2 = v_1 + v_5 ,\ c_3 = v_1 - v_5 ,\ c_4 = v_2 + v_6 ,\ c_5 = v_2 - v_6 ,\ c_6 = v_3 + v_7 ,\ c_7 = v_3 - v_7 .$$

Now we see that the elements $\{s_k\}$ can be calculated using only 24 additions. In matrix notation, the above calculations can be written more compactly as

$$\mathbf{S}_{8\times1} = \frac{1}{8}\mathbf{P}_8^{(5)}\mathbf{P}_8^{(4)}\mathbf{W}_8\mathbf{B}_{8\times1} \tag{8}$$

$$\mathbf{S}_{8\times1} = [s_0, s_1, s_2, s_3, s_4, s_5, s_6, s_7]^{\mathrm{T}} ,\ \mathbf{B}_{8\times1} = [b_0, b_1, b_2, b_3, b_4, b_5, b_6, b_7]^{\mathrm{T}} .$$

$$\mathbf{W}_8 = \mathbf{H}_2 \otimes \mathbf{I}_4 = \left[\begin{array}{cccc|cccc} 1 & & & & 1 & & & \\ & 1 & & & & 1 & & \\ & & 1 & & & & 1 & \\ & & & 1 & & & & 1 \\ \hline 1 & & & & -1 & & & \\ & 1 & & & & -1 & & \\ & & 1 & & & & -1 & \\ & & & 1 & & & & -1 \end{array}\right],$$

$$\mathbf{P}_8^{(4)} = \begin{bmatrix} 1 & & & & & & 1 & \\ & 1 & & 1 & & & & \\ & & 1 & & 1 & & & \\ & 1 & & -1 & & & & \\ & & -1 & & 1 & & & \\ & & & & & 1 & & 1 \\ 1 & & & & & & -1 & \\ & & & & & 1 & & -1 \end{bmatrix},\ \mathbf{P}_8^{(5)} = \begin{bmatrix} & 1 & 1 & & & & & \\ & -1 & 1 & & & & & \\ & & & 1 & 1 & & & \\ & & & -1 & 1 & & & \\ 1 & & & & & 1 & & \\ 1 & & & & & -1 & & \\ & & & & & & 1 & 1 \\ & & & & & & 1 & -1 \end{bmatrix}.$$

Fig. 4 shows a data flow diagram of the process for calculating the vector $\mathbf{S}_{8\times1}$ elements.

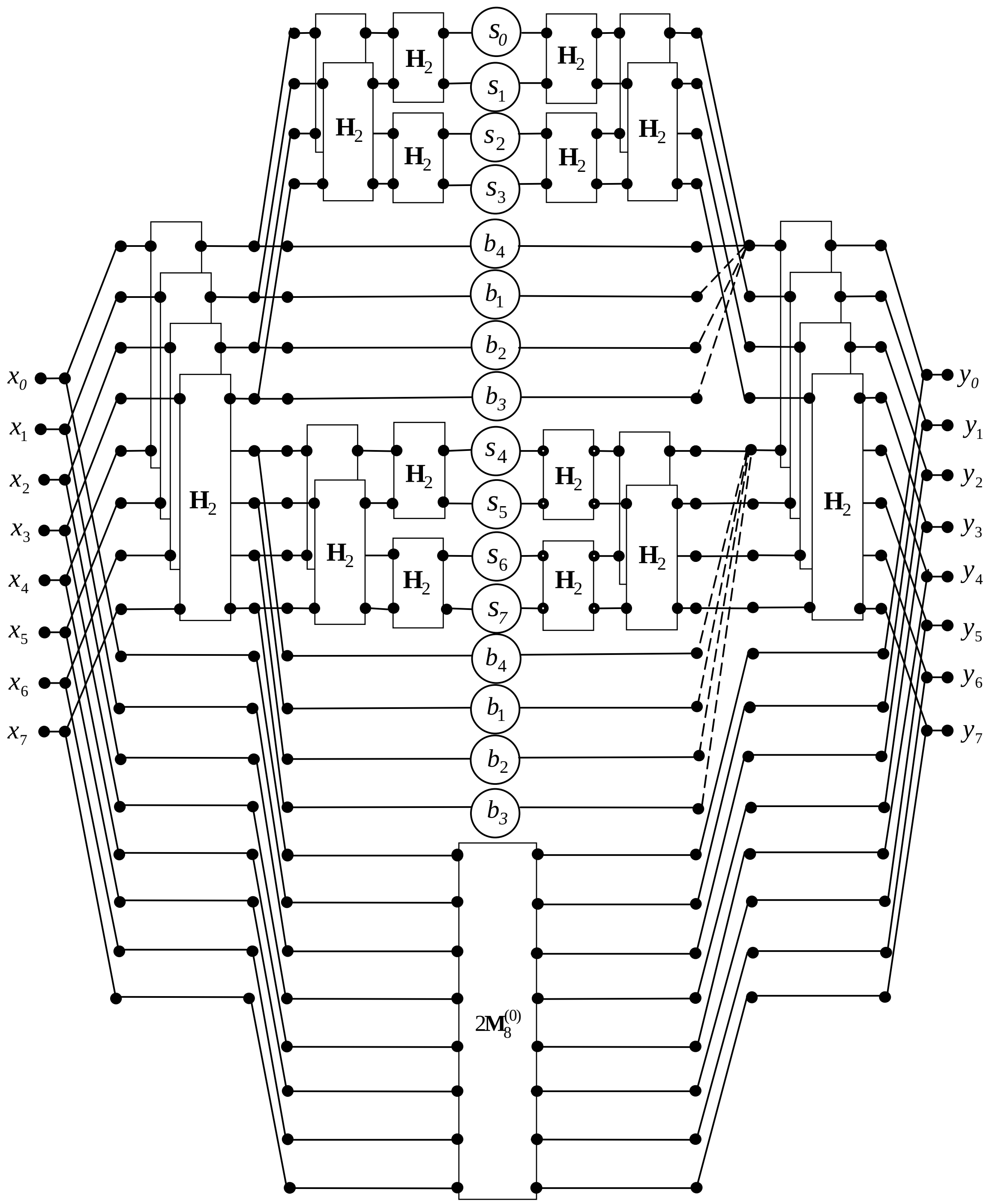

**Fig. 3.** Data flow diagram for rationalized split-octonion multiplication algorithm in accordance with the procedure (7).

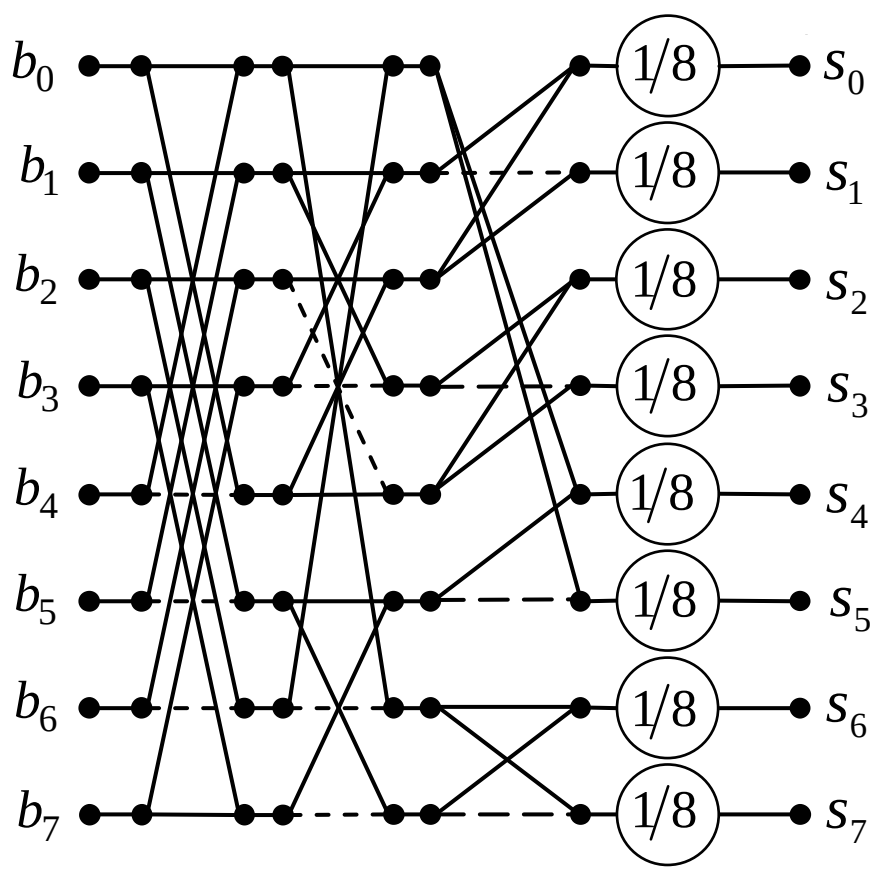

**Fig. 4.** Data flow diagram describing the process of calculating elements of the vector $\mathbf{S}_{8\times1}$ in accordance with the procedure (8).

## 4. Evaluation of computational complexity

We calculate how many real multiplications (excluding multiplications by power of two) and real additions are required for realization of the proposed algorithm, and compare it with the number of operations required for a direct evaluation of matrix-vector product in Eq. (2). Let us look to the data flow diagram in Figure 3. It is easy to verify that all the real multiplications which to be performed to computing the product of two split-octonions are realized only during multiplying a vector of data by the quasi-diagonal matrix $\mathbf{D}_{24}^{(2)}$. It can be argued that the multiplication of a vector by the matrix $\mathbf{D}_{24}^{(2)}$ requires 28 real multiplications and only a few trivial multiplications by the power of two. Multiplication by power of two may be implemented using convention arithmetic shift operations, which have simple realization and hence may be neglected during computational complexity estimation. So, the number of real multiplications required using the proposed algorithm is 28. Thus using the proposed algorithm the number of real multiplications to calculate the split-octonion product is significantly reduced.

Now we calculate the number of additions required in the implementation of the algorithm. It is easily to verify that the number of real additions required using our algorithm is 92.

## Conclusion

In this paper, we have presented an original algorithm that allows us to compute the product of two split-octonions with reduced multiplicative complexity. The proposed algorithm saves 36 real multiplications compared to the schoolbook algorithm. Unfortunately, the number of real additions in the proposed algorithm is somewhat greater than in the direct algorithm, but the total number of arithmetical operations is still the same. For applications where the "cost" of a real multiplication is greater than that of a real addition, the new algorithm is generally more efficient than direct method.